\documentclass[11pt,]{article}
\usepackage[]{graphicx}
\usepackage{makeidx}
\newcommand{\mathsym}[1]{{}}

\def\gsim{\mathrel{\raise.3ex\hbox{$>$\kern-.75em\lower1ex\hbox{$\sim$}}}}
\def\lsim{\mathrel{\raise.3ex\hbox{$<$\kern-.75em\lower1ex\hbox{$\sim$}}}}

\begin{document}

\begin{center}
{\huge {Cosmological bounds on the "millicharges" of mirror particles}}
\\ \vspace{1cm}
{Zurab Berezhiani} \footnote{zurab.berezhiani@aquila.infn.it, berezhiani@fe.infn.it}$^a$
{ and Angela Lepidi}\footnote{lepidi@fe.infn.it}$^b$
\\ \vspace{1cm} $^a$ {\it Universit\`a degli Studi di L'Aquila, 67010 Coppito, L'Aquila and
\\ INFN, Laboratori Nazionali del Gran Sasso, 67010 Assergi, L'Aquila, Italy}
\\ $^b$ {\it Universit\`a degli Studi di Ferrara, 44100 Ferrara and \\
Istituto Nazionale di Fisica Nucleare, Sezione di Ferrara, 44100 Ferrara, Italy}
\end{center}

\begin{abstract}
Mirror world, a parallel hidden sector with microphysics identical to
ordinary particle physics, can have several interesting phenomenological
and astrophysical implications and  mirror matter can be a natural candidate
for dark matter in the universe.
If the ordinary and the mirror photons have a kinetic mixing
due to the Lagrangian term $(\epsilon/2) F_{\mu\nu} F'^{\mu\nu}$,
then mirror particles effectively acquire the electric charges
$\sim \epsilon$ with respect to the ordinary photon, so that 
they become a sort of particles historically coined as "millicharged" 
though nowadays they must be called  more appropriately as "nanocharged". 
%If the kinetic mixing parameter is large enough (of order $10^{-9}$),
%the scattering of mirror baryons on ordinary matter can give a nice
%explaination of the DAMA/Libra observed annual modulation.
In this paper we revise the cosmological bounds on the kinetic mixing
parameter and in the case of exact mirror parity 
set an upper limit $\epsilon < 3 \times 10^{-10}$.  
Much weaker limit can be obtained  
in the case of asymmetric mirror sector, with 
an electroweak symmetry breaking scale larger than the ordinary 
electroweak scale. 
%and find very conservatively that this value is not compatible with cosmology.
\end{abstract}

%----------------------------------------------------------------------------------------------------------------------------------------
\section{\label{intro} Introduction}
%----------------------------------------------------------------------------------------------------------------------------------------

The old idea that there can exist a hidden mirror sector of particles
and interactions which is an exact duplicate of our visible world
\cite{Lee:1956qn} has attracted a significant interest over the last years.
The mirror theory is based on the product of two identical gauge factors
$G \times G'$ with an identical particle content.

The general procedure of doubling the gauge factors can be applied
to any gauge group, such as the standard model one
$[\mathcal{SU}(3) \times \mathcal{SU}(2) \times \mathcal{U}(1)]
\times [\mathcal{SU}(3)' \times \mathcal{SU}(2)' \times \mathcal{U}(1)']$
or the grand unified theories such as
$\mathcal{SU}(5) \times \mathcal{SU}(5)'$, etc.
(In the following, to distinguish between quantities referred
to the ordinary and to the mirror sector, the latter ones are marked with prime $'$).
Also, a "double" gauge factor naturally emerges in the
context of $E_8 \times E_8'$ superstring.

If the mirror world exists, universe should contain,
along with the ordinary particles: electrons, nucleons, photon, etc.
also their mirror partners: mirror electrons, mirror nucleons,
mirror photon, etc., with exactly the same mass spectrum and
interaction properties ({\it mirror parity}).
Any neutral ordinary particle, elementary or composite, can
have a mixing with its mirror counterpart exactly degenerate in mass.
E.g., photon can have
kinetic mixing with M-photon \cite{Holdom:1985ag,Glashow:1985ud,Carlson:1987si},
%which can be searched in the ortho-positronium oscillation into
%mirror ortho-positronium~\cite{ortho},
ordinary (active) neutrinos can mix with mirror (sterile)
neutrinos with interesting astrophysical implications
~\cite{Foot:1991py,Akhmedov:1992hh},
neutral $\pi$ mesons can mix with mirror $\pi'$ mesons,
neutrons with mirror neutrons \cite{Berezhiani:2005hv}, etc.\footnote{
In principle, ordinary and mirror sectors can have also different 
gravities. The ordinary to mirror graviton mixing and its 
cosmological implications were discussed in Ref. 
\cite{Berezhiani:2009kv}.
}
Such mixings can be
induced by the effective interactions between the O- and M-fields
mediated by some messengers, which may be some heavy gauge 
singlet particles, or heavy gauge bosons interacting with both sectors
~\cite{Berezhiani:1996ii}.

Mirror matter, being invisible in terms of ordinary photon and
interacting with ordinary matter only gravity, is a natural candidate
for dark matter (DM) consistent with cosmological tests
\cite{Berezhiani:2000gw,Ignatiev:2003js,Berezhiani:2003wj}.
In addition, the baryon asymmetry of the Universe can be generated
via out-of-equilibrium $B-L$ and $CP$ violating processes between
the visible and dark matter fractions in the universe \cite{Bento:2001rc}.
Such a mechanism can explain the closeness between  the baryonic
and dark matter fractions in the universe, providing naturally 
$\Omega'_{B}/\Omega_{B} \sim 1\div 5$ as far as the ordinary 
and mirror baryons have exactly the same masses 
\cite{Berezhiani:2003xm,Berezhiani:2006ac}.\footnote{
There is also a possibility that mirror parity is spontaneously
broken, and the electroweak symmetry breaking scale in 
mirror sector is  larger than the ordinary electroweak scale
\cite{Akhmedov:1992hh,Berezhiani:1995am}.
In this case mirror world would become a particular type
of a shadow world, with more heavy but 
less collisional and dissipative matter  
more resembling the cold dark matter (CDM),  
that we discuss in Section 3.}

Cosmological aspects, and in particular,
Big Bang Nucleosynthesis (BBN) bounds,
require that the temperature of mirror sector $T'$
should be smaller then the temperature $T$ of ordinary sector.
BBN is sensitive to the energy density of the universe at $T \sim 1$ MeV
\cite{Yao:2006px},
which is usually parametrized in terms of the
effective degrees of freedom 
$g_{\ast T} = g^{\rm st}_{\ast T} + \Delta g_{ \ast T}$ 
or the effective number of extra-neutrinos 
$\Delta N_\nu = \Delta g_{\ast T}/1.75$, where
$\Delta g_{\ast T}$ measures 
the contribution of any extra particle
species in addition to standard input $g^{\rm st}_{\ast T}=10.75$
as contributed by photons $\gamma$, electron-positrons $e,\bar e$
and three neutrino species $\nu_{e,\mu,\tau}$ at $T\sim 1$ MeV.
Therefore, the contribution of mirror photons $\gamma'$,
mirror electron-positrons $e',\bar e'$ and mirror neutrinos
$\nu'_{e,\mu,\tau}$ would correspond to
   \begin{equation}
    \label{N_nu_x^4}
    \left(\frac{\rho'}{\rho}\right)_\mathrm{BBN} =
    0.16 \, \Delta N_{\nu}  = x^4 \, ,
   %$ \Delta N_{\nu} \simeq 6.14 \, x^4 $,
    \end{equation}
or   $ \Delta N_{\nu} \simeq 6.14 \, x^4 $,
where $x = T'/T$ is a temperature ratio between two sectors
\cite{Berezhiani:2000gw}.
Hence, a conservative bound on the number of extra-neutrinos
$\Delta N_\nu < 0.5$ implies $(\rho'/\rho)_\mathrm{BBN} < 0.08$,
or $x < 0.5$.
A detailed analysis of the temporal evolution of the number
of degrees of freedom in both sectors can be found in \cite{Ciarcelluti:2008vs}.

The cosmological constraints from the CMB and large scale structure
of the Universe lead to more stringent limits, if 
one assumes that dark matter is entirely made of mirror baryons.
In this case, the perturbations in the mirror baryon fluid cannot grow 
before the mirror photon decoupling which occurs at the redshift 
$z'_{\rm dec} \simeq x^{-1}   z_{\rm dec}$ \cite{Berezhiani:2000gw}, 
where 
$z_{\rm dec} \approx 1100$ is the  redshift or the ordinary photon 
decoupling from the matter. 
However, for $x < 0.3$ the mirror photons decouple before 
the matter radiation epoch $z_{\rm eq} \approx 3000$ 
and so the density perturbations at the scales larger than 
the corresponding horizon size can undergo the linear growth. 
In this case, as it was shown in  \cite{Berezhiani:2003wj} 
via explicit computations, the linear power spectrum characterizing 
the large scale structures (LSS), 
at the scales $k/h < 0.2/$Mpc or so, 
as well as  the power spectrum of the CMB oscillations,  are 
practically indistinguishable from the standard CDM predictions. 
Somewhat stronger bounds emerge from the galaxy formation 
constraint. For example, by requiring that the density 
perturbations  corresponding to  
the galaxies like a Milky Way are not Silk-damped, 
we would get a bound $x < 0.2$, while   
the ``bottom-up" formation of smaller structures as dwarf galaxies 
as well as constraints from the Lyman-$\alpha$ forest would 
require $x < 0.1$ or so 
\cite{Berezhiani:2000gw,Ignatiev:2003js,Berezhiani:2003wj}. 
Nevertheless, in the following we take a more conservative 
limit $x < 0.3$  which is in fact a robust bound even in the case 
when mirror baryons constitute only a fraction of dark matter 
while the rest is provided by some kind of the CDM  
\cite{Berezhiani:2003wj},  
e.g. if $\Omega'_B = \Omega_B$  as in the limiting case 
implied  by the unified baryogenesis mechanism between 
the ordinary and mirror sectors \cite{Bento:2001rc}. 

The difference of the temperatures $T$ and $T'$ during 
the cosmological evolution can 
 occur if after inflation ordinary and mirror sectors are heated
at different temperatures; then they evolve adiabatically with the Universe
expansion, without strong first order phase transitions,
so that in both sectors the entropies are separately conserved.
Therefore, as far as there is no substantial
energy exchange between ordinary and
mirror sectors, the ratio $x=T'/T\simeq (s'/s)^{1/3}$ remains nearly
constant in time. 
Obviously, this is correct if during and after the inflation 
there is no significant entropy 
exchange between the ordinary and mirror sectors, 
which would be the case if they interact only via gravity.
However, if there are other particle processes between two sectors,
they should be  weak enough in order not to bring 
two sectors into thermal equilibrium with each-other.

One of the most interesting phenomena which may reveal the mirror sector
is the ordinary photon-mirror photon kinetic mixing, which arises when
the term $(\epsilon/2) F_{\mu\nu}  F'^{\mu\nu}$ is inserted in the
Lagrangian
\cite{Holdom:1985ag,Glashow:1985ud,Carlson:1987si}.
%Foot:1991bp,Foot:2000vy}.
This term is allowed by symmetries as far as the field strength tensors
of $\mathcal{U}(1)$ gauge bosons
$F_{\mu\nu} = \partial_\mu A_\nu-\partial_\nu A_\mu$, etc.,
are gauge invariants.
%When a kinetic mixing between the ordinary and the mirror photon
%is present  \cite{Holdom:1985ag,Carlson:1987si, Glashow:1985ud, Foot:1991bp},
Then the complete electromagnetic Lagrangian reads
    \begin{eqnarray}
    \label{lagrangian}
    \mathcal{L} = - \, \frac{1}{4}
    \Big(F_{\mu\nu}  F^{\mu\nu}   + F'_{\mu\nu}  F'^{\mu\nu}  +
    2 \epsilon F_{\mu\nu}  F'^{\mu\nu}\Big)
    - e \, q_f A_{\mu} (\bar f \gamma^{\mu} f) -
    e \, q_f A'_{\mu} (\bar f' \gamma^{\mu} f' )
    \end{eqnarray}
where $f$ and $f'$ stand for the charged particles as electrons or protons
respectively of ordinary and mirror sectors.
Performing the unitary transformation and the rescaling of the fields,
the kinetic terms can be diagonalized and canonically normalized
\cite{Holdom:1985ag,Foot:2000vy}.
One can choose a basis in which ordinary charged particles $f$
interact only with one combination $\tilde{A}_{\mu}$
("normal" photon), and
the other combination $\tilde{A}'_{\mu}$ of $A_{\mu}$ and $A'_{\mu}$,
a sort of paraphoton, interacts only with $f'$
while $f$s are "sterile" with respect to it.
Therefore, in this basis the
interaction term in the Lagrangian (\ref{lagrangian}) becomes:
    \begin{equation}
    \mathcal{L}_{int} =
    - \, e \, q_f \tilde{A_{\mu}} (\bar f \gamma^{\mu} f)
    - \, e' \, q_f \tilde{A'_{\mu}} (\bar f' \gamma^{\mu} f') -
    \epsilon \, e \, q_f \tilde{A_{\mu}} (\bar f' \gamma^{\mu} f') \, ,
    \end{equation}
so that charged mirror particles $f'$ electromagnetically interact
also with the normal photon, with the interaction constant
being suppressed by the kinetic mixing parameter $\epsilon$.
In other words, a mirror particle
with mirror electric charge $q_{f}$ acquires
also tiny ordinary electric charges $\epsilon q_f$.
For historical reasons such particles were called
{\it millicharged particles} (or MCPs).
From the point of view of the present cosmological
limits on $\epsilon$ are about $10^{-9}$,
it would be more proper to name them as possible
{\it nanocharged particles} (NCPs).

The photon-mirror photon kinetic mixing can induce ordinary-mirror positronium
oscillations and in principle can be detected via observing the invisible
decay channels of orto-positronium \cite{Glashow:1985ud,Gninenko:1994dr}.
The present experimental limits on the positronium decay imply
$\epsilon \leq 3 \times 10^{-7}$.
However sensibility of experiments can be improved to the level
$\epsilon \sim {\rm few}\times 10^{-9}$ \cite{Badertscher:2003rk}.

On the other hand, the nanocharged mirror nuclei, if they constitute
dark matter, could be detected by dark matter detectors.
In particular the results of DAMA/NaI \cite{Bernabei:1996vj}
experiment for dark matter detection could be nicely explained
by the scattering of nanocharged mirror nuclei if
$\epsilon \sim 10^{-9}$ \cite{Foot:2004ej}.

A few months ago the first results from the DAMA/Libra experiment
and the combined analysis with DAMA/NaI have been published
\cite{Bernabei:2008yi}. Both experiments performed a model-independent
and low-threshold dark matter (DM) search;
the recorded annual modulation signal has phase and periodicity
compatible with the dark matter expected signature.

Soon after, an interpretation of the DAMA results in terms of mirror
matter was proposed and compatibility with other experiments,
such as CDMS and XENON10, was analyzed \cite{Foot:2008nw}.
This interpretation is based on the idea that the signal detected
in DAMA may be due to scattering of nanocharged mirror nuclei
on ordinary matter. In particular, the best candidate for reproducing
the DAMA data and being unobservable in other experiments is the
nanocharged mirror oxygen. The analysis performed in \cite{Foot:2008nw}
shows that the interaction rate is proportional to
$(Z_{A'}\epsilon)^2 \xi_{A'}$ where $\xi_{A'}$ is the halo mass fraction
of the species $A'$, in our case mirror oxygen, that is
$\xi_{A'} = n_{A'} M_{A'} / (0.3$GeV/cm$^3)$.
In particular, the DAMA data can be reproduced if
$\epsilon \sqrt{\xi_{O'}} \sim 3 \times 10^{-10}$,
where $\xi_{O'}$ is the mirror oxygen mass fraction which is
typically assumed to vary between $10^{-3}$ and $10^{-1}$, that
implies $\epsilon \sim 10^{-8} \div 10^{-9}$.

However, if the mirror particles are nanocharged,
there are electromagnetic processes like
$e \bar e \leftrightarrow e' \bar e'$ leading
to energy transfer between the two sectors,
with the efficiency $\propto \epsilon ^2$.
Hence the mirror sector is heated and
the temperature ratio $x=T'/T$ increases.
In this way, the value of the kinetic mixing parameter
$\epsilon$ can be restricted by the cosmological bounds
on $x$.

%raise till the interaction freeze out point is reached.

The BBN constraints on the photon-mirror photon kinetic mixing
were discussed at first by Carlson and Glashow in 1987 and
the bound $\epsilon < 3 \times 10^{-8}$ was reported \cite{Carlson:1987si}.
This limit however needs to be updated in the light
of the modern data on the primordial element abundances.\footnote{
The generic NCPs,
without a specific reference to the mirror model,
have been worked out in Ref.
\cite{Davidson:1991si},
%Davidson:2000hf,Dubovsky:2003yn},
%Complete and more recent reviews can be found in \cite{}
%and \cite{Raffelt:1996wa},
where the bounds from accelerator
experiments, BBN, globular clusters, supernova 1987A,
white dwarfs and CMB were studied.
However, more attention was devoted to the light NCPs,
lighter than the electron, for which the astrophysical
bounds are more stringent.}

In this paper we revisit the cosmological bounds on $\epsilon$
or, in other words, bounds on the nanocharges of mirror particles,
that can be imposed by the analysis of the BBN and the CMB epochs.
Hence arises an important difference between the bound on $x$ coming
from BBN and the one from CMB: the first applies at
$T_\mathrm{BBN} \sim 1$ MeV, while the second applies at the matter
radiation equality epoch, when $T_\mathrm{CMB} \sim 1$ eV.
This difference must be taken into account when calculating bounds
on the model parameters, that is, the contributions to the mirror
energy must be calculated up to $T_\mathrm{CMB}$ when applying
the CMB limit on $x$.
We also study the case of spontaneously broken mirror parity,
in which case mirror particles, and in particular, mirror electron
becomes heavier than the ordinary ones, which significantly relaxes
the stringent bounds on the photon-mirror photon kinetic mixing
obtained for the case of exact mirror parity.

%----------------------------------------------------------------------------------------------------------------------------------------
\section{\label{bounds_calc} Bounds on the kinetic mixing parameter}
%----------------------------------------------------------------------------------------------------------------------------------------

If the kinetic mixing between the photons is present, there are
electromagnetic processes involving ordinary and mirror particles
and leading to energy and entropy exchanges between the two
sectors. At first order in the coupling constant $e$ there can be
pair annihilation and production $e \bar e \leftrightarrow e' \bar
e'$, elastic scatterings like $e e' \leftrightarrow e e'$ and
the plasmon decay $\gamma \rightarrow e' \bar e'$. For our
purposes, only the first process is significant. Indeed scattering
processes, which can take place only after mirror particles have
been created, lead to an energy transfer between the two sector
lower than that from the pair annihilation at least by a factor
$\sim x^3$. Plasmon effects, which generally give a dominant
contribution for the light MCP, with $m \ll m_e$
%\cite{Davidson:2000hf},
are ineffective for $m \geq m_e$ and therefore negligible for
us.\footnote{The plasmon decay becomes effective at $T\geq 10$ MeV
since the plasmon energy $\omega_P \sim 0.1T$ must be at least
$\sim 2m_e$. }
%Taking this process in account leads to a change in our
%bound of order $10^{-3}$.}

The amplitude of the annihilation process 
$e \bar e \rightarrow e' \bar e'$, 
for $m_e = m_{e'}=m$, 
is $\epsilon$ times  the $s$-channel amplitude 
for the process $e \bar e \rightarrow e \bar e$. 
The corresponding cross section 
%for the annihilation process   
reads
    \begin{eqnarray}
    \label{good}
    \sigma_\epsilon
    %= \epsilon^2 \sigma
    = \epsilon^2 \, \frac{4\pi \alpha^2}{3} \, \frac{(s +
    2m^2)^2}{s^3}\, ,
    \end{eqnarray}
To calculate the energy exchanges between the two sectors we need
the interaction rate $\Gamma_\epsilon$, which is defined in terms
of the average the cross section times the velocity of colliding
particles, $\Gamma_\epsilon \equiv \langle \sigma_\epsilon v
\rangle n$. For relativistic electrons, $T > m$, $\Gamma$ has the
form
%\cite{Davidson:2000hf}:
%
    \begin{eqnarray}
    \label{s_G_scat_urev}
    \Gamma_\epsilon = \epsilon^2 \Gamma_1, \quad \quad
    \Gamma_1 = 0.2 \, \alpha^2 T
    % \frac{\zeta_3}{2 \pi}
    \end{eqnarray}
%
%Comparing $\Gamma$
which should be compared with the Hubble parameter
%$H$, that is $\Gamma \epsilon^2 < H$ at the
%neutron-proton freeze out ($T_W\sim 0.8$ MeV). The Hubble parameter $H$ is
%
    \begin{eqnarray}
    \label{H_def_T}
    H %\equiv \frac{\dot R}{R}
    = 1.66 \,g_{\ast T}^{1/2} \frac{T^2}{M_P}  \, ,
    \end{eqnarray}
$M_P$ being the Planck mass and $g_{\ast T}$ the total number
of degrees of freedom at the temperature $T$. At the BBN epoch, $T
= 0.8$ MeV, we have $g_{\ast T} \simeq 10$ and hence we see that
$\Gamma_\epsilon < H$ is satisfied
%.
%Hence, the condition $\Gamma \epsilon^2 \leq H$ at $T_W = 0.8$ MeV
%leads to a very rough bound on $\epsilon$, that is
if $\epsilon < 5 \times 10^{-9}$ or so \cite{Lepidi:2008hb}.

However,  this is only an estimate and for deriving
more precise bounds
on $\epsilon$ more accurate calculations are needed, by solving
corresponding Boltzmann equations and taking
%most accurate because we solve directly
%a differential equation in the mirror sector energy and we take
into account the low energy tale (below $1$ MeV) of
the cross section of the processes $e \bar e \rightarrow e' \bar e'$.
The latter should be treated more precisely
for setting the BBN bounds on $\epsilon$,
and more importantly, for discussing the process
at the lower temperatures, $T \ll 1$ MeV, since the
corresponding asymptotic value of $T'/T$ is relevant for
the cosmological features of the mirror dark matter
related to the mirror photon decoupling and the growth of primordial
perturbations.
When $T < 1$ MeV the relativistic approximation is no longer valid,
so we cannot use $\Gamma$ in eq. (\ref{s_G_scat_urev}).
The thermal average at low temperature, when $T \leq 3m_e$,
can be calculated and it has the form
\cite{Gondolo:1990dk}:
    \begin{eqnarray}
    \label{Gondolo_Gelmini_Moller}
    \langle \sigma_\epsilon v \rangle =
    \frac{1}{8m_e^4 T K_2^2(m_e/T)}
    \int_{4m_e^2}^{\infty} \sigma_\epsilon  \cdot (s-4m_e^2) \sqrt{s}
    K_1\left( \frac{\sqrt{s}}{T} \right) \;ds
    \end{eqnarray}
where $K_1$ and $K_2$ are the modified Bessel functions of the
second kind,
%AAA
%{\bf $\sigma$ is a generic cross section}
%AAA
 and $v$ is the M\"oller velocity.

In the following, we neglect the energy loses for the mirror sector
and thus take that its energy density rescales
as $\rho \propto g_{\ast T} T^4$.
%The energy density of the ordinary sector obeys:
%
 %   \begin{eqnarray}
 %   \label{rho_diff_eq_ord}
 %   \frac{d \rho}{dT} - 4 \frac{\rho}{T} = 0 %f_\rho (T)
%   -\frac{0.6 \, M_P}{\sqrt{g}} \times  \frac{\Gamma(T) \rho_e(T)}{T^3}
%   \equiv - f_\rho (T),
 %   \end{eqnarray}
%
%where the source term is absent because the energy transfer from the mirror
%to the ordinary sector is negligible.
We assume also that
energy transferred  to the mirror sector is conserved, i.e.
% in the mirror sector, that is
%
    \begin{eqnarray}
    \label{energy_transfer}
    \frac{d}{dt} (\rho' R^3) + p' \frac{d}{dt} (R^3) =  
    \Gamma_\epsilon R^3 n_e \langle E \rangle \, ,
    \end{eqnarray}
where $\langle E \rangle$ is the average energy transferred to
the mirror sector per an $e\bar e \to e' \bar e'$ process.
Then by excluding the scale factor $R$ and
substituting $n_e \langle E \rangle$ approximately by
$\rho_e$ in a source term, we obtain:\footnote{This approximation
of the exact Boltzmann equations leads to more conservative
limits on $\epsilon$ as far as it underestimates the amount of
the transferred energy. }
%recalling that $p'=\rho'/3$
%
    \begin{eqnarray}
    \label{energy_transf}
    \frac{d \rho' }{dt} + 3H (\rho' + p') =   \Gamma_\epsilon \rho_e
    %\langle E \rangle
    \end{eqnarray}
where $\rho'$ and $p'$ are respectively the total energy density
and the total pressure of the mirror sector: 
$p'\approx \rho'/3$ as far as the relativistic component 
is dominant. 
%(photons $\gamma'$)
%dominate over the contribution of mirror electron-positrons.
% and $ -  \Gamma_\epsilon R^3
%n_e \langle E \rangle$ represents the source term leading to
%energy exchange between the sectors.
In our equation we only
consider the energy transfer from ordinary to mirror
sector without backreaction because the mirror energy density is
smaller than the ordinary one by approximately a factor $x^4 \ll 1$
and hence the energy transfer from mirror to ordinary
sector is negligible. 
The electron number density $n_e$ which enters
in $\Gamma_\epsilon$ (\ref{s_G_scat_urev}) and 
the energy density $\rho_e$ 
are taken at their equilibrium values at
the temperature $T$:
%and the same can be done for the energy density $\rho$:
%
    \begin{eqnarray}
    \label{number_density}
    n_e(T) = %\int f(\mathbf{k}) d^3\mathbf{k} =
    \frac{2}{\pi^2}
    \int_{m_e}^{\infty} dE \, \frac{\sqrt{E^2-m_e^2} \:E}{\exp(E/T) + 1} \, ,
    \quad \quad
    \rho_e(T) = %\int E f(\mathbf{k}) d^3\mathbf{k} =
    \frac{2}{\pi^2}
    \int_{m_e}^{\infty} dE \, \frac{\sqrt{E^2-m_e^2} \:E^2}{\exp(E/T) + 1}
    \end{eqnarray}
%
%where $g_A$ is the spin degeneracy of the particle species and $+$ and $-$
%stand respectively for fermions and bosons.
%Eq. (\ref{Gondolo_Gelmini_Moller})
%can be more conveniently expressed in terms of $T$ by using eq. (\ref{H_def_T})
% together with
Substituting $t =  0.3 \, M_P / g_{\ast T}^{1/2} T^2$,
%Hence,
the above equation can be rewritten as
    \begin{eqnarray}
    \label{rho_diff_eq}
    \frac{d \rho'}{dT} - 4 \frac{\rho'}{T} = - \epsilon^2
    %f_\rho (T)
    \frac{0.6 \, M_P}{\sqrt{g_{\ast T}}} \times  \frac{ \Gamma_1(T) \rho_e(T)}{T^3}
    \equiv - \epsilon^2 f_1 (T)
    \end{eqnarray}
and its solution
%of eq.(\ref{rho_diff_eq})
can be presented as
    \begin{eqnarray}
    \label{rho_sol}
    \label{rho_eq_gen}
    \frac{\rho'(T)}{\rho(T)} = \epsilon^2  Q_T, \quad \quad
    Q_T = -
    \frac{30}{\pi^2 g_{\ast T}} \int_{\infty}^T dy  (f_1(y)/y^4)
    \end{eqnarray}
where we assume that the energy of the mirror sector was
negligible with respect to the ordinary one at the beginning.
This is the most conservative initial condition: indeed if
we assume that the energy of the mirror sector was comparable
with the one of the ordinary sector, the bounds on $\epsilon$
become even more stringent.

    \begin{figure}[htbp]
    \begin{center}
     \includegraphics[scale=1]{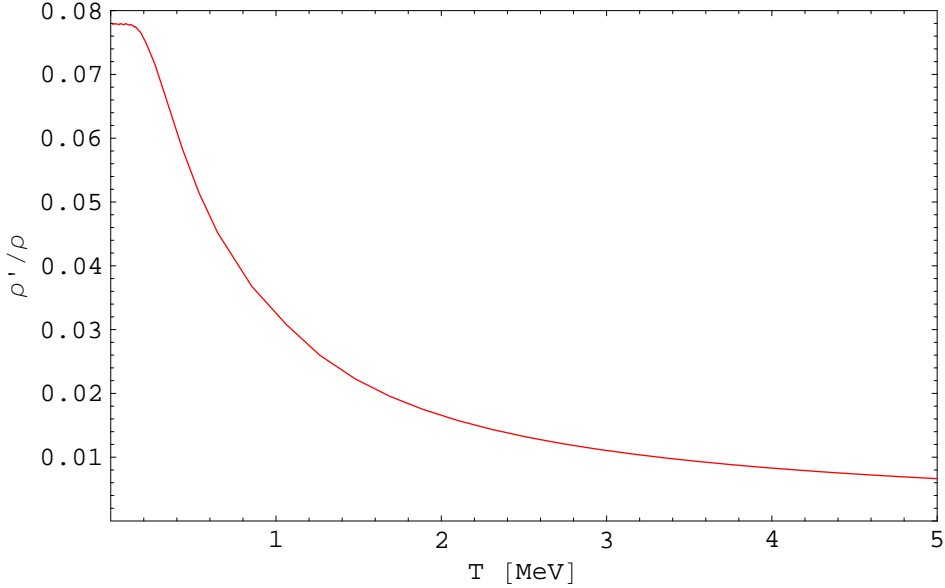}
     %{Bound_x_03.eps}
    \caption{
    The parameter $Q_T= \epsilon^{-2} (\rho'/\rho)$
    %$\epsilon$
    in units of $10^{18}$. 
    It corresponds to the value of $\rho'/\rho$ for $\epsilon = 10^{-9}$. 
    %for $x = 0.3$. The upper region in the plot is excluded by cosmology.
    }
    \label{plot_03}
    \end{center}
    \end{figure}

The BBN bound on $\epsilon$ can be obtained 
solving this equation numerically and imposing that
$\rho' / \rho = \epsilon^2 Q_T < 0.16 \, \Delta N_\nu$ 
at $T \simeq 0.8$ MeV. 
On the other hand, for determining the cosmological 
bounds from the LSS and CMB,  $x < 0.3$, one has 
to integrate eq. (\ref{rho_sol}) till the temperatures 
$T \ll 1$ MeV. 
Our results for a function $Q_T$ are shown in Fig. \ref{plot_03}. 
We can see that below $T \simeq 0.2$ MeV,
$Q_T$ does not change anymore and 
it goes asymptotically to $Q_0 \approx 8\times 10^{16}$. 
Physically this is due to Boltzmann 
suppression of the electron and positron densities 
 below $T \simeq m_e$,
that leads to strong suppression of the energy
transfer from the ordinary to the mirror sector.
As we see, the value of $Q_T$ at $T=0.8$ MeV, 
relevant for the BBN epoch, is smaller 
roughly by a factor 2 than the asymptotic value $Q_0$.  

Therefore, at the BBN epoch, $T \simeq 0.8$ MeV,
we obtain 
%from (\ref{N_nu_x^4}) 
the following bound:
%$\Delta N_\nu < 0.5$ or according to , $(\rho'/\rho)_\mathrm{BBN} < 0.08$
\begin{equation}
    \label{BBN-bound}
    %\mathrm{BBN}:  \quad \quad
    \epsilon  < \sqrt{\frac{(\rho'/\rho)_\mathrm{BBN} }{Q_{T=0.8\, {\rm MeV}}}} \approx 
    % 1.7 \times 10^{-9} \left(\frac{x_\mathrm{BBN}}{0.5} \right)^2 =
    1.5 \times 10^{-9}
    \left(\sqrt{\frac{\Delta N_\nu}{0.5}}\right)^{1/2}
    %_\mathrm{BBN}
    \end{equation}

The cosmological bound $x < 0.3$ or so concerns the
temperature ratio $x=T'/T$ rather then the ratio of the densities
$\rho'/\rho$. Taking that at $T \ll 1$ MeV we have
$\rho(T) \propto g_{\ast T} T^4$, with $g_{\ast T}\simeq 3$
that apart of the photons takes into account also the contribution
of neutrinos
decoupled from the thermal bath at the temperatures $T > 2$ MeV,
while $\rho'(T) \propto g'_{\ast T'} T'^4$, with $g'_{\ast T'}= 2$
as contributed only by mirror photons since the mirror neutrinos cannot
be produced at lower temperatures, we get the limit
    \begin{equation}
    \label{bound}
    \epsilon  <  \sqrt{ \frac{g'_{\ast T'}/g_{\ast T} } {Q_0} } x^2
    <  3  \times 10^{-10}
    \left( \frac{x} {0.3}\right)^2
    %_{\rm M-R\,\, equality}
    \end{equation}
Thus, a conservative cosmological bound requiring that
$T'/T < 0.3$ at the Matter-Radiation equality epoch gives
$\epsilon < 3 \times 10^{-10}$, while the galaxy formation bound
$x < 0.2$ would give 
$\epsilon < 2 \times 10^{-10}$
about twice stronger limit.\footnote{
Let us recall that this bound is valid only if dark matter
is entirely constituted by mirror baryons.}
The interpretation \cite{Foot:2008nw} of the DAMA/Libra results
in terms of Rutherford scattering of mirror baryons
on ordinary matter is hardly compatible with these
cosmological bounds on $\epsilon$,
since then it requires the mass fraction of mirror oxygen $\xi_{O'}$
of about $0.6$, which seems too much.
The primordial chemical composition in mirror sector is not
the same as in ordinary one \cite{Lepidi:2008hb} and thus
also the present element abundances are presumably different.
So a detailed analysis of the stellar evolution may be performed
(see e.g. in \cite{Ciarcelluti:2004ip})
to calculate what should be the present concentration of oxygen.
Nevertheless it seems hard to obtain such a high value.
Finally, we stress that assuming $x \sim 0.25$ leads to
$\xi_{O'} \sim 1$, that is, the mirror sector should be exclusively
made of oxygen.

%----------------------------------------------------------------------------------------------------------------------------------------
\section{\label{asym_mir} The asymmetric mirror model and dark matter}
%----------------------------------------------------------------------------------------------------------------------------------------

Cosmological observations indicate that the present universe is nearly flat,
with the total energy density $\rho_\mathrm{tot}$ very close to the critical
$\rho_\mathrm{c}$:
$\Omega_\mathrm{tot} \equiv \frac{\rho_\mathrm{tot}}{\rho_\mathrm{c}}
\simeq 1$.
Non-relativistic matter in the universe consists of a baryonic (B) and a
dark (DM) component where $\Omega_\mathrm{DM} \sim 0.21$ and
$\Omega_\mathrm{B} \sim 0.04$ \cite{Yao:2006px}.
The relationship $\Omega_\mathrm{DM} / \Omega_\mathrm{B} \sim 5$ is
puzzling ({\it fine tuning problem}) since both $\rho_B$ and $\rho_{DM}$
scale as $a^{-3}$ during the universe expansion and thus their ratio is
independent on time. So a priori there is no apparent reason by which
they should be so close to each other.

An answer to this problem is  naturally found in the mirror sector
physics, in particular if we assume that the mirror parity is broken,
as it was suggested in \cite{Akhmedov:1992hh,Berezhiani:1995am}.
Indeed, we stated in the
introduction that the mirror parity implies that the mass spectrum and
the interaction properties are the same in the two sectors. Nevertheless,
if mirror parity is spontaneuosly broken, the electroweak symmetry
breaking scales are different in the ordinary and in the mirror sector
and this would lead to different physics in the two sectors.
In particular, if we call the Higgs expectation values in the ordinary
and in the mirror sector respectively $\langle \phi \rangle = v$ and
$\langle \phi' \rangle = v'$, we can define the parameter
$\zeta = v'/v$ and see immediately that the mass spectrum of 
elementary fermions
$f$ (leptons or quarks) and gauge bosons $W,Z$ 
changes according to
$m_{f'}/m_{f} = \zeta $ and $M_{W', Z'}/ M_{W, Z} = \zeta  $
\cite{Berezhiani:1995am}.
At the same time the $\Lambda_\mathrm{QCD}$ constant scales according
to $\Lambda'/ \Lambda = \zeta^{0.28}$ \cite{Berezhiani:2000gh}.
Hence, if we assume e.g. $\zeta \sim 100$, 
the mirror electron mass scales up to 
$m'_e \sim 50$ MeV while  the (composite) masses 
of mirror nucleons  become approximately
$M'_{B} \sim 5$ GeV that account for the scaling 
both of $\Lambda' \sim 3.5 \Lambda$ and of the 
light quark masses $m'_{u,d} \sim 10^2 m_{u,d}$ 
\cite{Berezhiani:2006ac}.

Let us now analyze what are the cosmological bounds on the kinetic
mixing parameter $\epsilon$ in the asymmetric mirror scenario. The
scattering $e^+e^- \leftrightarrow e'^+ e'^-$ is again the only
relevant process and has a threshold at energy of order
$T_\mathrm{thr} \sim m_{e'} = \zeta m_e$  
when ordinary electrons are still relativistic. 
Hence we can use the relativistic
approximation for $\Gamma$ in eq.(\ref{s_G_scat_urev}) and solve
analytically eq.(\ref{rho_sol}), which gives
    \begin{eqnarray}
    \label{}
    \frac{\rho'(T)}{\rho(T)} =
    7.7 \times 10^{-7} \epsilon^2 \, \frac{M_P}{\zeta m_e}
    \leq 0.16 \Delta N_\nu
%   \frac{\left[ \int_{\infty}^T (f_\rho(y)/y^4) dy \right]}{\pi^2 g_\mathrm{tot} / 30}
    \end{eqnarray}
that implies 
$\epsilon \leq 2 \times 10^{-8} \sqrt{\zeta \Delta N_\nu/50}$.
In the asymmetric case the bound on $x$ from CMB does not apply: 
as far as $m'_e \gg m_e$, mirror photons decouple 
much before the matter radiation equality epoch even if two sectors 
have the same temperatures, and thus 
mirror dark matter should behave practically as a cold dark matter, as 
far as the large scale structure and the CMB oscillations are 
concerned.  
Thus the only constraint comes from the BBN limit 
on the number of extra-neutrinos which we conservatively 
take as 
$\Delta N_\nu \leq 0.5$ \cite{Berezhiani:1995am}, 
which e.g. 
for $\zeta = 100$ or $m'_e = 50$ MeV,
transforms in the bound $\epsilon \leq 2 \times 10^{-8}$.

Let us recall, that according to ref. \cite{Foot:2008nw}, 
elastic scattering of a mirror nucleus  mediated by 
photon-mirror photon kinetic mixing gives best fit to 
the DAMA annual modulation when its mass is of about 16 GeV. 
In the case of exact mirror parity, when the ordinary and mirror 
nucleons are exactly degenerate in mass, the proper 
mirror nucleus would be the oxygen, having 
$M_{O'} \sim 16$ GeV and atomic number $Z_{O'} = 8$. 
On the other hand, in the case of asymmetric mirror (shadow) 
sector, with $v'/v \sim 10^2$, when mirror nucleons become 
about 4-5 times heavier than their ordinary brothers, 
the best candidate would be mirror  helium, with mass
$M_{He'} \sim 16$ GeV and $Z_{He'} = 2$.
Since the interaction rate in DAMA is proportional to
$(Z \epsilon)^2 \xi$, we need $\epsilon$ about to be
$4$ times higher to compensate the charge difference
between mirror helium and oxygen, that is
$\epsilon \sim 4 \times 10^{-9}$, which is compatible with
our cosmological bound.

Nevertheless, it should be considered that in the asymmetric
sector the mass difference between light quarks scales as
$(m'_d - m'_u) \sim \zeta (m_d - m_u)$ and consequently the
mass difference between the mirror neutron and proton is some
hundred MeV, while 
$\Lambda'_\mathrm{QCD} \sim \zeta^{0.3} \Lambda_\mathrm{QCD}$.
Such a large mass difference cannot be compensated by
the nuclear binding energy. Hence in the asymmetric mirror model
also the neutrons bounded in nuclei may be unstable against 
$\beta$ decay \cite{Berezhiani:1995am} 
and thus heavy nuclei may be not formed.
Obviously, in this case mirror helium will not exist as a stable nucleus, 
%in the case If the neutron is not stabilized inside the  helium-like configuration, 
and the only possible candidate for dark matter 
can be the mirror hydrogen, with mass of about 4-5 GeV, 
which still can be appropriate for the DAMA/LIBRA signals, 
but  the fit is much worse. 
In the supersymmetric extensions, if the parameters characterizing 
the up-down Higgs VEV ratios are not equal between two sectors, 
if $\tan\beta'  > \tan\beta$, 
than there is also possibility that the neutron rather than proton is 
the stable baryon in the mirror sector. In this case no mirror 
electrons and protons will be present in the present universe  
while dark matter will be due to mirror neutrons, 
and so  practically no interesting limit can be settled  
for the photon-mirror photon kinetic mixing. 
However, in this case the latter mixing cannot be at work for 
the dark matter direct detection.

%----------------------------------------------------------------------------------------------------------------------------------------
\section{\label{conclusions}Conclusion}
%----------------------------------------------------------------------------------------------------------------------------------------

We have discussed cosmological implications of the parallel mirror world
with the same microphysics as the ordinary one but having smaller
temperature and the photon kinetically mixed with the ordinary one.
 In this model charged mirror particles acquire small electric charges
 (nanocharges) proportional to the mixing parameter $\epsilon$.

In particular if mirror baryons are nanocharged,  
$\epsilon\sim 10^{-9}$, the 
 scattering of mirror nuclei on the standard matter
 may produce the annual modulations observed by 
 the DAMA/LIBRA experiment 
 and be at the same time avoid the (un)detection limits 
 from other experiments looking for dark matter, such as CDMS
 and XENON 10 \cite{Foot:2008nw}.
Actually the interaction rate does not depend simply on $\epsilon$,
but also on the mass fraction of oxygen $\xi_{O'}$ and on its charge $Z_{O'} =8$ and is proportional to $(\epsilon Z)^2 \xi_{O'}$. Since $\epsilon\sim 10^{-9}$ corresponds to $\xi_{O'} \sim 0.1$, smaller values of $\epsilon$ can be allowed if the amount of oxygen is higher than $0.1$.

In this paper we have studied in detail the energy transfer from the ordinary to the mirror sector in order to calculate cosmological bounds on the kinetic mixing parameter $\epsilon$ at the BBN and CMB epochs.
When the mirror electrons are at least as heavy as the ordinary ones the pair annihilation of ordinary $e \bar e$ in $e' \bar e'$ is the only process by which there is a relevant energy transfer. We integrated the cross section at low energy in order to take into account the energy transfer below $T \sim 1$ MeV, which should be considered when imposing bounds from CMB.
Our most conservative result is $\epsilon \leq 3 \times 10^{-10}$, which may be compatible with DAMA assuming $\xi_{O'} \sim 1$, which seems however an unnatural composition.

The cosmological bound on $\epsilon$ was calculated also in the asymmetric mirror model, where all particles are heavier than their ordinary partners \cite{Akhmedov:1992hh}. Under this hypothesis the most conservative bound is $\epsilon \leq 2 \times 10^{-8}$, corresponding to $0.5$ extra-neutrinos at the BBN epoch, while the more stringent cosmological 
bound  coming from the LSS and CMB pattern does not apply in this case. 
In the asymmetric mirror model the best candidate to fit DAMA is helium, 
with $Z'=2$, which requires a value of $\epsilon$ compatible with 
the above limit. 
A problem can however arise: the light quarks mass difference scales as the ratio of the Higgs VEVs in the two sectors, $\zeta=v'/v \sim 100$ and so does the neutron-proton mass difference, while $\Lambda_\mathrm{QCD}$ scales as $\zeta^{0.3}$ \cite{Berezhiani:2000gh}. Hence the nuclear binding energy may be not high enough to make bound neutrons stable. 
This problems should be further investigated in future researches.

%----------------------------------------------------------------------------------------------------------------------------------------
\paragraph{Note added} 
%----------------------------------------------------------------------------------------------------------------------------------------
In the case of exact mirror parity, our limit on $\epsilon$ is somewhat 
stronger than the one obtained in the recent work \cite{Ciarcelluti:2008qk}. 

%----------------------------------------------------------------------------------------------------------------------------------------
\paragraph{Acknowledgements}
%----------------------------------------------------------------------------------------------------------------------------------------
The work is supported in part by the European FP6 Network "UniverseNet"
MRTN-CT-2006-035863.

%----------------------------------------------------------------------------------------------------------------------------------------

\end{document}